\documentclass[aps,prd,twocolumn,nofootinbib,showpacs,preprintnumbers]{revtex4}
\usepackage[dvips]{graphicx}
\usepackage{amssymb}
\usepackage{verbatim}
\usepackage{psfrag}
\def \beq{\begin{equation}}
\def \eeq{\end{equation}}
\def \bea{\begin{eqnarray}}
\def \eea{\end{eqnarray}}
\def \md{m_{\scriptscriptstyle D}}
\def \ssp{\sigma_{\rm s}}

\def \Tr {\rm Tr}
\def \tc{T_c}
\def \qq{\bar{q} q}

\def \bet0{\beta_0}
\def \bet1{\beta_1}
\def \simgt{\,\rlap{\lower 7.5 pt\hbox{$\mathchar \sim$}}\raise 3 pt \hbox{$>$}\,}
\def \simlt{\,\rlap{\lower 7.5 pt\hbox{$\mathchar \sim$}}\raise 3 pt \hbox{$<$}\,}

\def\lsim{\raise0.3ex\hbox{$<$\kern-0.75em\raise-1.1ex\hbox{$\sim$}}}
\def\gsim{\raise0.3ex\hbox{$>$\kern-0.75em\raise-1.1ex\hbox{$\sim$}}}
 
\def \tc{T_c}

\begin{document}

\title{The Spatial String Tension and Dimensional Reduction in QCD}
\author{M. Cheng$^{\rm a}$, S. Datta$^{\rm b}$, 
J. van der Heide$^{\rm c}$, K. Huebner$^{\rm d}$, F. Karsch$^{\rm c,d}$, O. Kaczmarek$^{\rm c}$,\\ 
E. Laermann$^{\rm c}$, J. Liddle$^{\rm c}$, R. D. Mawhinney$^{\rm a}$, 
C. Miao$^{\rm c}$, P. Petreczky$^{\rm d,e}$, \\ K. Petrov$^{\rm f}$,
C. Schmidt$^{\rm d}$, W. Soeldner$^{\rm d}$ and T. Umeda$^{\rm g}$
}
\affiliation{
$^{\rm a}$Physics Department,Columbia University, New York, NY 10027, USA\\
$^{\rm b}$Department of Theoretical Physics, Tata Institute of Fundamental
Research, Homi Bhabha Road, Mumbai 400005, India\\
$^{\rm c}$Fakult\"at f\"ur Physik, Universit\"at Bielefeld, D-33615 Bielefeld,
Germany\\
$^{\rm d}$Physics Department, Brookhaven National Laboratory, 
Upton, NY 11973, USA \\
$^{\rm e}$RIKEN-BNL Research Center, Brookhaven National Laboratory, 
Upton, NY 11973, USA \\
$^{\rm f}$DESY, Platanenallee 6, Zeuthen, Germany
$^{\rm g}$Graduate School of Pure and Applied Sciences, University of Tsukuba,
Tsukuba, Ibaraki 305-8571, Japan\\
}

\begin{abstract}
We calculate the spatial string tension in (2+1) flavor QCD
with physical strange quark mass and almost physical light
quark masses using lattices with temporal extent $N_{\tau}=4,~6$ and
$8$. We compare our results on the spatial string tension with 
predictions of dimensionally reduced QCD. This suggests that 
also in the presence of light dynamical quarks dimensional reduction works 
well down to temperatures $1.5T_c$.
\end{abstract}
\preprint{BI-TP 2008/12, BNL-NT-08/21 , TIFR/TH/23}
\maketitle

\section{Introduction}
At high temperature, strongly interacting matter undergoes a
transition from a confining, chiral symmetry broken phase to a phase
which is chirally symmetric and deconfining in nature. For pure gluonic
matter as well as for QCD with massless quarks this transition is known to be
a genuine phase transition, 
while for QCD with physical quark masses it is most
likely only a smooth (though rapid) crossover \cite{milc04,fodor,tc06}. 
In both cases, thermodynamic observables like energy and entropy density
show a sharp increase at the transition point, but 
approach the expected behavior of an asymptotically free quark-gluon gas 
only at very high temperatures. 

The phase just above $\tc$ is more complicated 
than a weakly interacting gas of quarks and gluons. Non-perturbative features 
like the
(Debye) screening of electric modes as well as the possible appearance
of a mass gap in the magnetic sector give rise to  well-known problems that
make a straightforward perturbative analysis of the high temperature phase
of QCD fail. While infrared divergences arising at a
momentum scale $p \sim g T$ can still be taken into account by 
resummation \cite{Zhai,Arnold,spt97,Blaizot,Andersen}, 
the softer modes at scale $p \sim g^2 T$ reflect genuine non-perturbative 
physics and cannot be handled by any resummation \cite{linde}. 

The non-perturbative physics arising from the magnetic sector of QCD 
shows up at distances that are large on the scale of the inverse
temperature, $R\sim 1/(g^2 T)$. It finds its most prominent reflection in 
the survival of an area law behavior for space-like Wilson loops, {\it i.e.}
confinement of magnetic modes. The spatial string tension, $\sigma_s$, 
extracted from the area law behavior of spatial Wilson loops, has been studied 
early on in lattice calculations performed in the pure gauge sector of 
QCD \cite{detar,polonyi}. Detailed studies of the temperature dependence
of $\sigma_s$ in SU(2) and SU(3) gauge theories 
\cite{bali93,bielefeld,eos} 
showed the expected dependence on the magnetic scale $\sqrt{\sigma_s} \sim
g^2(T)T$ and established a close relation to the string tension of a
3-dimensional gauge theory.  
A recent detailed comparison of $\sqrt{\sigma_s}$, calculated 
in lattice regularized QCD, with calculations performed in the 
framework of dimensionally reduced QCD
\cite{mikko} was extremely successful in describing the temperature dependence
of $\sqrt{\sigma_s}$ quantitatively even at temperatures close to the 
transition temperature. 

Dimensional reduction makes very specific
predictions for modifications induced in the magnetic sector of QCD through
the presence of fermion degrees of freedom, quarks. In fact, modifications
are expected to be small and the dominant structure of e.g. the spatial
string tension is still expected to be controlled by a purely gluonic, 
3-dimensional SU(3) gauge theory.
A possible source for failure has, however, 
been discussed in the past \cite{gavai}; dynamical quarks, more precisely 
quark anti-quark pairs, can give rise to additional light bosonic
quasi-particle modes that are not described by a 3-dimensional gauge theory.
This makes it interesting to check to what extent dimensional reduction
also is capable of describing physics in the magnetic sector of QCD in the
presence of light dynamical quarks. 

In this paper we calculate the spatial string tension in QCD with light
dynamical quarks. For the first time we cover in such an analysis a large
temperature range and perform calculations at three values of the lattice
cut-off. 
We will present a detailed comparison of the lattice calculations
with predictions based on dimensional reduction. Some preliminary results 
of this work have been presented in \cite{frithjof}. 

This paper is organized as follows. In the next section we give a brief 
discussion of the concepts of dimensional reduction as far as they are relevant
for this work and the analysis of the spatial string tension. In Section III
we present details of our calculation of spatial pseudo-potentials and 
in Section IV we 
discuss the determination of the spatial string from them. 
A comparison of these results
with dimensional reduction is given in Section V. Finally we give our 
conclusions in Section VI.

\section{Dimensional reduction}

The different non-perturbative length scales of ${\cal O}(1/gT)$ and
${\cal O}(1/g^2T)$, respectively, that one encounters in the perturbative
analysis of QCD at high temperature are convenient for the
reformulation of QCD in terms of a hierarchy of effective field theories
that clearly separate the physics at these different scales \cite{rob}. 
This allows the disentanglement of sectors that can be treated 
in perturbation 
theory from sectors that are only accessible to non-perturbative techniques. 

Since the partition
function of an equilibrium field theory can be written as an Euclidean
field theory with the time direction being a torus of extent $1/T$, the 
temporal modes of fields are proportional to the Matsubara frequencies,
$2 \pi n T$, where $n$ is an integer
for bosonic and a half-integer for fermionic fields. At large
temperatures, therefore, the non-static bosonic fields and the 
fermionic fields become very heavy and can be integrated out to get an 
effective theory for distance scales\footnote{Distances
are naturally measured  
in units of $1/T$, and $R T \sim 1$ 
is a natural unit since the average spatial separation of partons in the 
high temperature phase is $\sim 1/T$.}
$RT \gg 1$ which involves the static modes 
only \cite{rob}.
This gives an effective 3-dimensional theory\footnote{In 
this normalization convention the spatial gauge fields have 
dimension one, while the electric field $A_0$ has a dimension $1/2$.},
\bea
S_3^E &=& \int d^3x \, {1 \over g_E^2} \,
\Tr \, F_{ij}({\bf x}) \, F_{ij}({\bf x}) + \, \Tr \, [D_i,A_0({\bf x})]^2
\nonumber \\
&{}&  
\, + \, \md^2 \Tr A_0({\bf x})^2 + \lambda_A (\Tr A_0({\bf x})^2)^2 \; ,
\label{eq.dm1} 
\eea
with the static mode of the 
temporal component of the gauge field, $A_0$, appearing as 
an adjoint scalar field. The parameters of this effective theory can be
calculated in perturbation theory to any order \cite{braaten95,kajantie97}.  
At leading order one has 
$g_E^2=g^2 T, \lambda_A=(9-n_f) g^4 T/(24 \pi^2)$ 
and $m_D^2 = g^2 \, T^2 \, (1 + n_f / 6)$
for $n_f$ massless quark flavors. 
The deconfined phase of high temperature QCD corresponds to the
symmetric (confining) phase of the above 3d adjoint Higgs theory 
(see discussion in \cite{kajantie97,karsch98}). The confining nature of this 
theory is the origin of the infrared problem of high temperature QCD 
for momenta $p \sim g^2 T$.  At very
high temperatures also the Debye mass will be large, $m_D\gg g^2 T$, 
and the heavy $A_0$ field can thus also be integrated out.
Therefore, for physics of distance scales $R \gg {1 / gT}$, the $A_0$ 
fields can be integrated out which leads to a pure three-dimensional gauge
theory, 
\beq
S_3 = \int d^3x \, {1 \over g_3^2} \, \Tr \, F_{ij}({\bf x}) \, F_{ij}({\bf x})
\;.
\label{eq.dm2} \eeq
The gauge coupling $g_3^2$ can be calculated in terms of $g_E$ and
$\md$. This has been done to 2-loop accuracy \cite{giovannangeli}. 
At leading order  $g_3^2 = g_E^2$.
The hierarchy of
effective theories is referred to as the dimensional reduction scheme.
The use of the dimensional reduction scheme in combination
with lattice calculations 
performed within the effective 3-dimensional theory allows the treatment of the infrared problem of perturbative QCD at high temperature
and the consistent extension of the weak coupling expansion of thermodynamic quantities to higher orders, for instance the
$O(g^6)$ contribution to the thermodynamic potential ( pressure ) of QCD \cite{kajantie03}.

In pure SU(N) gauge theory, the underlying nonperturbative structure of
the high temperature theory can directly be seen in calculations of spatial 
Wilson loops. The static $\qq$ energy in the finite temperature theory 
is given by Wilson loops in space-time planes, $W(R,\tau)$, which are 
expected to satisfy a perimeter law above $\tc$, thus signaling
deconfinement. In contrast, entirely space-like Wilson loops, which
are restricted to fixed 
time hyperplanes, show an area law, as they can be
mapped to the $\qq$ energy in the confining 3-dimensional theory 
\cite{detar,polonyi} as indicated above. 
Detailed studies of spatial Wilson loops in SU(2) and SU(3)
gauge theories have
corroborated the dimensional reduction picture 
\cite{bali93,bielefeld,eos}.  
The spatial string tension $\ssp$, the coefficient of the area term,
\beq
\ssp = - \lim_{R_1,R_2 \to \infty} {1 \over R_1 R_2} \ln W(R_1,R_2),
\label{eq.ssp} \eeq
has been found to be in very good quantitative agreement with that of
the dimensionally reduced theory. Since $g_3^2$ sets the mass scale in
the 3-dimensional gauge theory, one has $\sqrt{\sigma_3} = c g_3^2$ where $c$
is a numerical factor. Dimensional reduction therefore predicts 
\beq
\sqrt{\sigma_s} = c g^2(T) T
\label{eq.lo} \eeq
to leading order, where $g^2(T)$ is the temperature-dependent running
coupling of the four dimensional theory and $c$ is a constant that can be
determined through a non-perturbative (Monte Carlo) calculation of Wilson loops
in a 3-dimensional SU(3) gauge theory
\cite{bielefeld,teper}.
Already this tree level matching was found to be valid to a good
accuracy at temperatures as low as $2 \tc$ \cite{bielefeld,eos}.  
This is in accordance with studies of other correlation functions 
in SU(2) and SU(3) gauge
theories, which also show dimensional reduction to be a valid
approximation already at temperatures $T \simeq 2 T_c$. 
It is worth mentioning that the perturbative separation of electric and
magnetic scales, $\md \sim gT > g^2T$, is not valid at
these temperatures, where $g(T)>1$ and where the electric gluons thus are 
the lightest degrees of freedom \cite{prd,peter,hart}. 
Strictly speaking,
the step leading from Eq.~(\ref{eq.dm1}) to Eq.~(\ref{eq.dm2}) thus 
should not be taken.
However, the Higgs sector does not seem 
to have a significant effect on the correlation functions 
of purely gluonic 3d gauge fields \cite{peter,hart}.

The situation may be somewhat different in QCD. While the quarks 
acquire an effective mass of ${\cal O}(\pi T)$, they make a qualitative
difference in the long distance physics: since they provide a source for
screening of fundamental charge, one expects that at sufficiently 
large distances 
$RT \gg {2 \pi / g^4}$ (spatial) string breaking takes place, 
which cannot be described by the effective theories $S_3^E$ or $S_3$,
respectively, as
those do not include fundamental charges. Of course, the distance 
(in units of $1/T$) one needs to reach to observe this deviation increases with
temperature, so it remains an important question to ask how well
the (spatial) string tension of the four dimensional theory, 
extracted at large distances
but before string breaking sets in, can be explained by the
dimensionally  reduced theory. In principle it is possible
to write down the theory which includes also the lowest fermion Matsubara
frequency \cite{vepsalainen}. 
This version of the dimensionally
reduced theory would be able to describe the string breaking as
well. 
At distances smaller than the string breaking scale, 
however, and given the weak dependence of the 3d gauge fields on
the dynamics of the adjoint scalar fields discussed above we 
would expect that the presence of fermionic modes with mass of
${\cal O}(\pi T)$ will have very little influence on 
Wilson loops, and the spatial string tension extracted from them.
On the other hand,
from a comparison of screening masses in a theory 
with $n_f$ = 4 degenerate light flavors, Ref. \cite{gavai} argued 
that the presence of dynamical quarks could impede  dimensional reduction. 
Thus, the question of an effective description 
of spatial Wilson loops
in terms of a dimensionally
reduced theory at not-too-large temperatures 
is not settled and needs to be treated more extensively for QCD.

\section{Spatial potential at finite temperature}
We have carried out a
study of the spatial Wilson loops for QCD in the temperature 
interval [$0.9\;\tc, 4\; \tc$]. In this study we fixed the strange quark
mass ($m_s$) to its physical value, and the $u,d$ quarks
are taken to be degenerate with mass $0.1 m_s$. These light quark masses are 
about a factor two heavier than physical $u,d$ quarks, and give a pion 
mass of about $220$ MeV. 
The gauge configurations we analyze here were generated for the study of 
the equation of state of hot strongly interacting  matter \cite{eos_paper}. 
The gauge action
uses Wilson plaquette and rectangle terms. For the fermion action the
p4fat3 staggered discretization \cite{p4fat3} was used, which adds 
a bended three-link term to the standard staggered action, and the 
1-link gauge connections are smeared with the 3-link staples. 
In our analysis we have used lattices with temporal extent 
$N_{\tau}=4,~6$ and $8$ and aspect ratio $N_{\sigma}/N_{\tau}\ge 4$.
The lattice spacing has been set using the Sommer parameter $r_0$ \cite{sommer}.
When quoting the results in physical units the value $r_0=0.469$~fm
has been used \cite{gray}.
Given the above temporal lattice sizes and the temperature interval 
covered in our numerical
calculations, the range of lattice spacings varied from $0.3$fm
to $0.05$fm. Further details on the generation of gauge configurations
and the action used in these calculations can be found in 
Ref. \cite{eos_paper}.

To interpret the spatial Wilson loops as the exponential of the static $\qq$
potential in the effective 3-dimensional theory, we use a transfer
matrix formalism in the z direction: the loops are constructed as
$W(R,Z)$, where $R$ is the distance between the static quark and
anti-quark in the xy-plane. The loops in a timeslice were constructed
using the Bresenham algorithm \cite{bresenham} and they were averaged over 
the timeslices. The static $\qq$ potential, for a transfer matrix 
acting in the z direction, is defined as \cite{creutz}
\beq
a V_s(R) = - \lim_{Z \to \infty} \ln {W(R, Z+1) \over W(R, Z)} \; .
\label{eq.wloop} \eeq

To obtain a better signal-to-noise ratio it is important to be able to
reach the asymptotic limit already for moderate values of $Z$. To achieve
this, the gauge fields connecting quark and anti-quark were smeared 
using 3-link staples perpendicular to the direction of the gauge field.
The smearing procedure has been applied iteratively. We found that
for small values of the gauge coupling $\beta$ ($\beta<3.41$) 10 APE smearing 
steps with a smearing coefficient of 0.4 were sufficient to give the 
asymptotic value already at $Z$ = 3. For larger $\beta$ we use 30 to 60 
smearing steps.

The Wilson loop ratios in Eq. (\ref{eq.wloop}) were performed for different
$Z$, and a plateau is reached by $Z$=3. To avoid any remaining contamination
from higher states, the logarithm of the ratio of  Wilson loops was fitted to the ansatz 
$- a V_s+b \cdot \exp(-cZ)$. The results from the direct ratio at $Z=3$ generally agree well 
with the value obtained from fits.
We used in our analysis estimates for the string tension coming from
Eq.~(\ref{eq.wloop}) with $Z=3$ as well as results obtained from an
extrapolation to  $Z = \infty$.  Differences in these results 
have been treated as a systematic error in our determination of $\sigma_s$.

At each value of the temperature we have generated several thousand 
(up to 33,000) gauge field configurations using a Rational Hybrid Monte-Carlo 
(RHMC) algorithm. The total number of gauge configurations generated at each
gauge coupling can be found in Ref. \cite{eos_paper}.
The calculation of spatial Wilson loops has been performed on
gauge configurations separated by 10 RHMC trajectories. 
To reduce autocorrelation, the measurements were bunched 
into blocks and errors on $aV_s(R)$ were calculated
using Jackknife samples. By varying the block-size of the Jackknife samples
we checked that the calculated errors do not change significantly.
\begin{figure}[htb]
\includegraphics[width=9cm]{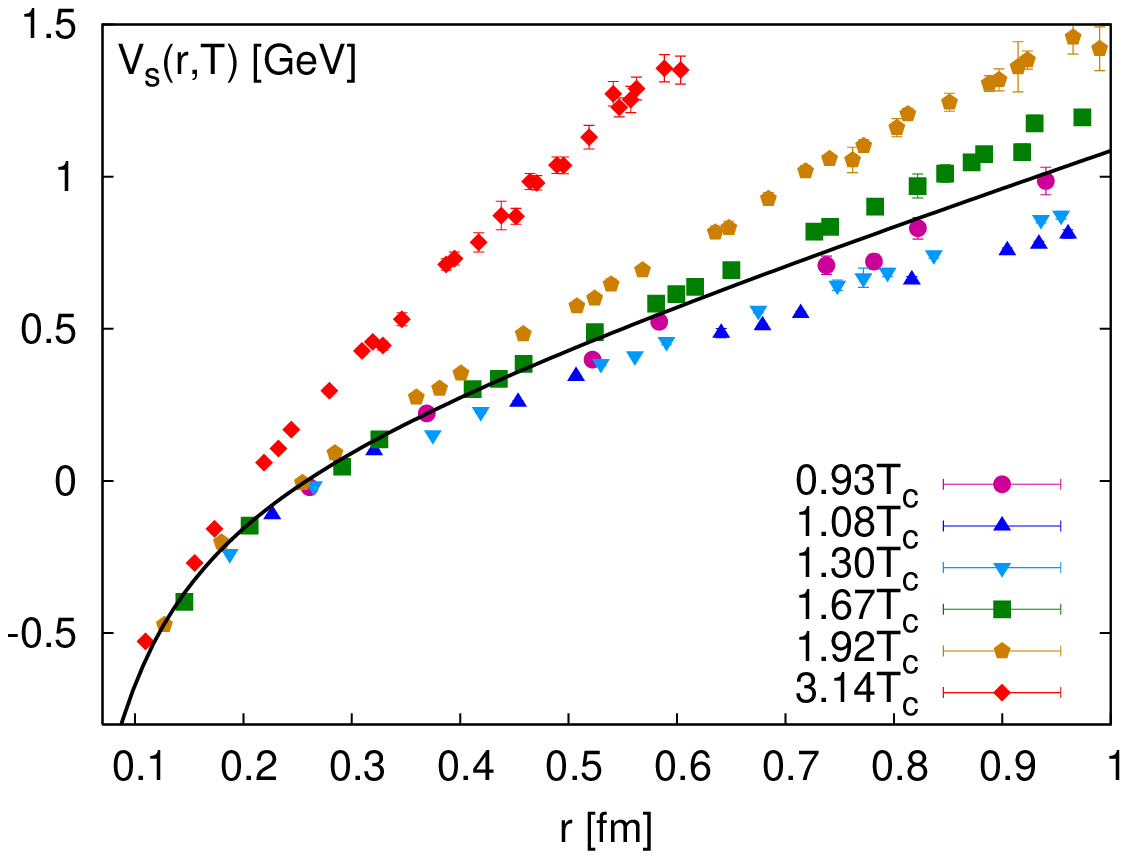}
\includegraphics[width=9cm]{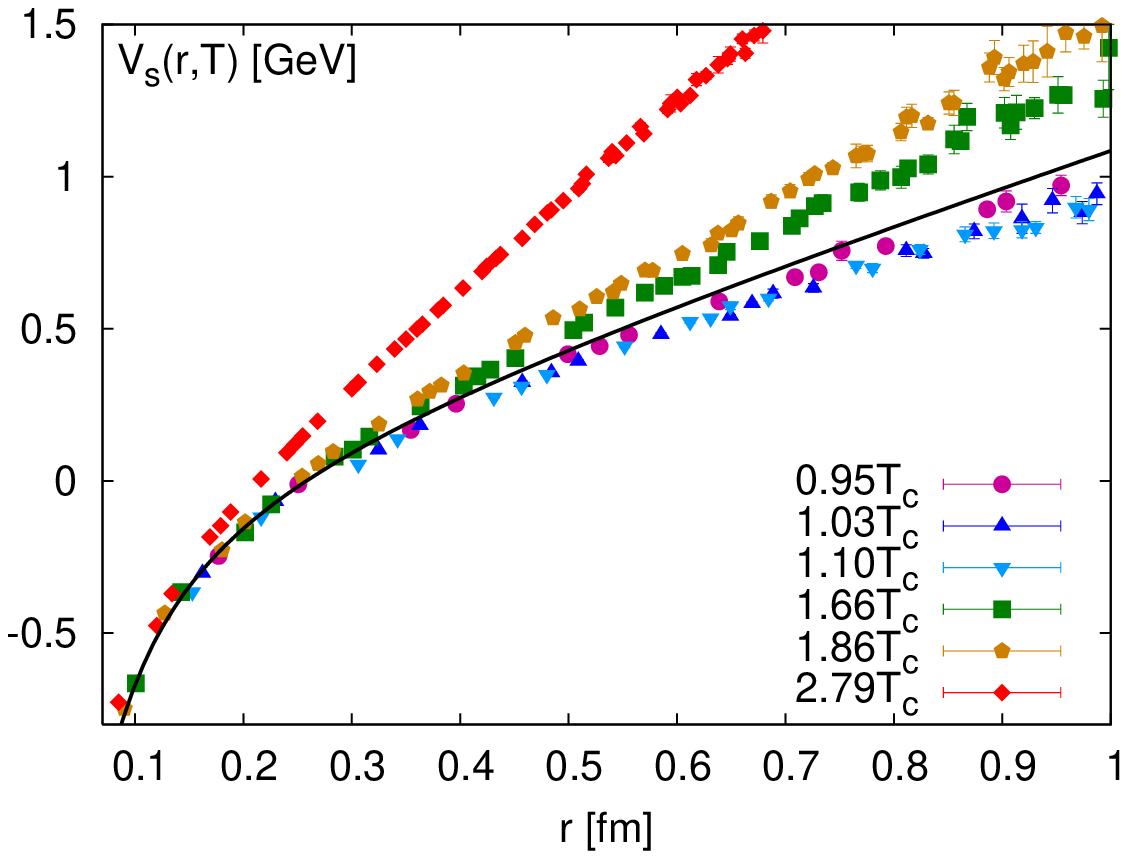}
\caption{
The spatial potential $V_s(R,T)$ calculated on $N_{\tau}=4$ (top)
and $N_{\tau}=6$ (bottom) lattices. The reduced temperature $T/T_c$
has been calculated using the values of $r_0/a$ determined in 
\cite{eos_paper} as well as the values $r_0 T_c=0.484$ for $N_{\tau}=4$ 
\cite{tc06} and $r_0 T_c=0.466$ for $N_{\tau}=6$ \cite{eos_paper}.
The line  is the fit to zero temperature potential calculated at $\beta=3.63$.
}
\label{fig:pot}
\end{figure}

Let us first discuss the most prominent features of the 
spatial potential at finite temperatures and compare it to
the usual zero temperature potential.
Note that for the zero temperature lattices, of course, timelike and
spatial Wilson loops are identical. 
In Figure~\ref{fig:pot} we show the spatial 
potential calculated for $N_{\tau}=4$ and $N_{\tau}=6$ at a few selected 
values of the temperature. We have subtracted the renormalization constant
determined at zero temperature in Ref.~\cite{eos_paper} to make the
comparison between different temperatures easier.  
Figure \ref{fig:pot} shows that below
$T_c$ the spatial Wilson loop shows very little change with
temperature, while above the transition temperature a strong
temperature dependence is seen. Only at very short distances  the 
spatial potential remains, 
within errors,
%
temperature independent also 
above $T_c$. For temperatures close to the transition temperature the
spatial potential falls below the zero temperature potential. The effect is
more pronounced for $N_{\tau}=4$ than for $N_\tau = 6$ which might
indicate that this is a cut-off effect. As the temperature increases 
the slope of the spatial potential at large distances, {\it i.e.}
the spatial string tension, clearly increases in agreement with
observations made previously in pure gauge theories. This will be discussed
in more detail in the next section.

We find no evidence of string breaking occurring in the high temperature
spatial potentials. This, however, may not be too surprising as it is 
known from studies of the heavy quark potential at zero temperature that
smeared Wilson loops at numerically accessible perimeters
are not well suited
for studying string breaking.
Polyakov loop correlation functions have been found to perform 
much better in this
respect \cite{deTar}. 
Therefore one might want to also analyze spatial 
Polyakov loop
correlation functions to get more sensitive to string breaking effects in
spatial potentials.

\section{Spatial string tension}
In order to extract the spatial string tension from the large
distance behavior of the spatial pseudo-potentials we have  to
rely on fits. However, the choice of an appropriate  fit Ansatz is less 
obvious than at zero temperature.
At zero and low temperatures the (pseudo-) potential is 
usually well described by
\begin{equation}
V_s(R) = -\frac{\alpha}{R} + c_0 + \sigma_s R,
\label{eq:potfit}
\end{equation}

The $1/R$ term, in $3+1$ dimensions, may arise from two
contributions, from Coulombic
perturbative gluon exchange at small distances, 
approximately $\lsim 0.2$ fm,
and from string fluctuations relevant at distances 
$\gsim 0.3$ fm. The later is often referred to as the L\"uscher term.
In fact, in the proportionality factor of the string fluctuation
term, $\alpha_L$, is a universal constant,
\begin{equation}
\alpha_L = \frac{(D-2)\pi}{24}  \; ,
\end{equation}

that only depends on the space-time dimensionality $D$ \cite{Luescher}.
As dimensional reduction manifests itself as temperature increases
one should expect that the contribution arising from string fluctuations
should gradually turn from $\pi / 12$ to $\pi / 24$.
In fact, finite temperature corrections to the string fluctuation
term have been calculated previously \cite{deForcrand}
and have been found to be of relevance for the analysis of the temperature
dependence of the (conventional) heavy quark potential \cite{Kaczmarek}. 
At the same time and for the same reason, the Coulombic 
term due to perturbative gluon exchange is expected to change
from a $1/R$ behavior appropriate in 4 dimensions to  
a logarithmic $R$ dependence in an effectively 3 dimensional theory.
It is this gradual change in the short and intermediate distance
part of the pseudo-potential that is responsible for 
ambiguities in the fit Ansatz. We will discuss in the following our
strategy to deal with this ambiguity.

From our data, we could not clearly disentangle a ${\rm log}(R)$ term
from a $1/R$ behavior. In fact, addition of the ${\rm log}(R)$ term to 
the form of Eq. (\ref{eq:potfit}) only makes the fit very noisy.
Similarly we could not restrict the fit intervals to such large
distances where $1/R$ terms of whatever origin, Coulombic or
string fluctuations, could be safely neglected.
We therefore have fitted our data to Eq.(\ref{eq:potfit})
with 3 free parameters. 
We note that this is not too severe a restriction as we are not
interested in a detailed description of the $R$-dependence of the 
pseudo-potentials
but rather want to get control over its large distance behavior.

\begin{figure}[htb]
\includegraphics[width=9cm]{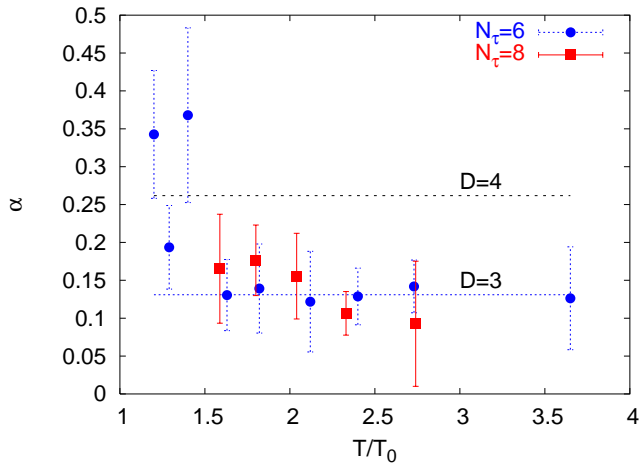}
\caption{Fitted values for $\alpha$ compared with the  coefficients of the L\"uscher term 
for 3 and 4 dimensions, $N_\tau = 6,~8$. 
The temperature has been scaled by $T_0=200 MeV$, where $r_0 T_0=0.47619$.}
\label{fig:fittedalpha}
\end{figure}

On our finer lattices ($N_{\tau}=6$ and $8$ ) we  observe that the coefficient of the $1/R$ term, 
{\it i.e.} $\alpha$,
decreases with increasing temperature
and approaches $\pi/24$ at high temperatures.  This is shown
in Figure~\ref{fig:fittedalpha}.
The fit results for $\alpha$ are somewhat dependent on the fit range.
We have adopted to choose a minimum separation $R_{min}$ between
$0.7$ and $1$ times $r_0$.
At sufficiently large distances the Coulombic term will not 
be seen, with only string fluctuations contributing.  
Fixing the value of $\alpha$ to the coefficient of the L\"uscher term, $\alpha_L$,
may describe the potential at large distances.  
This Ansatz was fitted to the potential at distances greater than $r_0$.  
For small temperatures, $T<1.5T_c$, 
with $\alpha$ fixed to $\pi/12$
the values of $\sigma$ coming from these 2 parameter fits are systematically 
larger than those coming from the 3 parameter fits.  At high temperatures, 
however, the spatial string tension obtained from two parameter fits with 
$\alpha=\pi/24$ and three parameter fits agree within statistical errors.

\begin{figure}
\includegraphics[width=9cm]{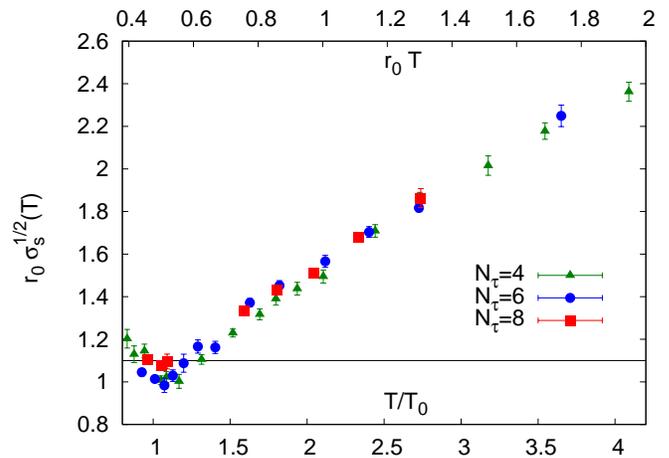}
\caption{The string tension in units of $r_0$ as function of the temperature
calculated on $N_{\tau}=4,~6$ and $8$ lattices. 
For better visualization 
the temperature axis has been scaled by $T_0=200MeV$ corresponding 
to $r_0 T_0=0.47619$.}
\label{fig:sigr02}
\end{figure}
The values of the spatial string tension 
for different $N_{\tau}$ are summarized in Table \ref{tab}. 
For $N_{\tau}=6$ and $8$  the string tension has been determined from
3 parameter fits. For the $N_{\tau}=4$ case we used mostly two parameter fits with
different values of $\alpha$. The errors on the spatial string
tension shown in the table are predominantly systematic due to the choice of the
fit form and fit interval. The temperature dependence of the spatial 
string tension is shown
in Figure \ref{fig:sigr02}.

For temperatures $T<1.2T_c$ the spatial string tension 
is close to the zero temperature string tension and is linearly rising with 
temperature for $T > 1.5T_c$ This is expected if dimensional reduction holds
at these temperature values. We will discuss this in more detail in the next 
section. It is possible that also the rapid drop in the coefficient $\alpha$
of the Coulomb term for $T>1.5T_c$ is related to the onset of the 3-dimensional physics reflected in the linear
rise of the spatial string tension in this temperature region. 
\begin{table*}
\begin{tabular}{|lll|lll|lll|}
\hline
\multicolumn{3}{|c|}{$N_{\tau}=4$} &  \multicolumn{3}{|c|}{ $N_{\tau}=6$}  &  \multicolumn{3}{|c|}{ $N_{\tau}=8$} \\
\hline
~~$\beta$  & $~~r_0 T$ & ~~$\sigma a^2$ & ~~$\beta$  & ~~$r_0 T$ & ~~$\sigma a^2$ &  ~~$\beta$  & ~~$r_0 T$  & ~~$\sigma a^2$  \\
\hline
3.150     & 0.367(18) & 0.535(36)       & 3.43       & 0.441(1) &  0.156(5)        &  3.53       &  0.458(4) & 0.0905(7)      \\
3.210     & 0.396(9)   & 0.577(33)       & 3.46       & 0.482(3) &  0.123(3)        &  3.57       &  0.501(3) & 0.0722(11)        \\
3.240     & 0.417(8)   & 0.458(27)       & 3.47       & 0.511(3) &  0.103(7)        &  3.585     &  0.520(13) & 0.0695(29)      \\
3.277     & 0.449(5)   & 0.406(21)       & 3.49       & 0.537(5) &  0.102(5)        &  3.76       &  0.756(8)  & 0.0486(12)    \\
3.335     & 0.500(3)   & 0.254 (10)      & 3.51       & 0.571(10)&  0.101(7)        &  3.82       &  0.858(12)  & 0.0435(6)      \\
3.351     & 0.517(3)   & 0.244 (12)      & 3.54       & 0.615(6) &  0.100(5)        &  3.92       &  0.974(9)  & 0.0375(6)        \\
3.382     & 0.556(3)   & 0.203(13)       & 3.57       & 0.668(4) &  0.084(4)        &  4.00       & 1.110(10) & 0.0357(6)    \\
3.410     & 0.626(5)   & 0.195(8)         & 3.63       & 0.775(7) &  0.087(2)        &  4.08       & 1.304(28) & 0.0319(7)      \\
3.460     & 0.723(4)   & 0.181(5)         & 3.69       & 0.867(8) &  0.078(2)        &               &                &                         \\
3.490     & 0.806(8)   & 0.167(6)         & 3.76       & 1.008(10)&  0.067(2)        &               &                &                         \\
3.510     & 0.856(15) & 0.165(4)         & 3.82       & 1.144(15)&  0.0610(6)        &               &                &                         \\
3.540    & 0.922(9)    & 0.152(6)         & 3.92       & 1.298(11) &  0.0544(5)     &               &                &                         \\
3.570    & 1.002(7)    & 0.139(5)         & 4.08       & 1.738(37) &  0.0465(7)     &               &                &                         \\ 
3.630    &1.163(10)   & 0.135(5)         &               &               &                      &                &                &                         \\ 
3.690    &1.300(12)   & 0.126(6)         &               &               &                      &                &                &                         \\ 
3.760    &1.513(15)   & 0.111(5)         &               &               &                      &                &                &                         \\ 
3.820    & 1.688(24)  & 0.104(3)         &               &               &                      &                 &                &                         \\ 
3.920    & 1.947(17)  & 0.094(2)          &              &               &                      &                 &                &                         \\ 
\hline
\end{tabular}
\caption{The spatial string tension in lattice units extracted from Coulomb plus linear fits on $N_{\tau}=4,~6$ and $8$ lattices (see text for the details).
The temperature scale, $r_0 T$, is obtained from the values of $r_0/a$ 
calculated in Ref. \cite{eos_paper}.
}
\label{tab}
\end{table*}

\section{Comparison with the prediction of dimensionally reduced theory}

As has been discussed in Section II we expect that at high temperatures the 
spatial string tension should be given by 
$\sqrt{\sigma_s}=c \cdot g_E^2(T)$. At 
temperatures several orders of magnitude larger than the transition
temperature, when the static electric field $A_0$ can be integrated out,
the effective theory is just a pure SU(3) gauge theory. 
The proportionality 
coefficient  $c$ is just a constant and has been determined to 
be \cite{teper} $c=0.5530(10)$,
corroborating an earlier value of $0.554(4)$ \cite{bielefeld}.
In the interesting temperature range of a few times $T_c$, however, $c$ will 
depend on the mass $m_D$ and coupling
$\lambda_A$ of the $A_0$ field and thus on the temperature. In the light of the
approximate decoupling of the 3d scalar and gauge fields we expect that the
dependence of the coefficient $c$ on these parameters should be weak and its
value should be close to the pure gauge value given above. Indeed,    
the calculations of the string tension in 3-dimensional adjoint Higgs models 
show only a weak
dependence on the parameters of the scalar sector and its value is close to the
pure gauge value \cite{hart}. 
Unfortunately, the statistical accuracy of the spatial string tension
calculated in the 3d adjoint Higgs model is significantly lower than in the
pure gauge case and no continuum extrapolation has been performed. For the relevant
case of three quark flavors and temperatures of about $2T_c$ the value of $c$ is
about $2\%$ lower than the pure gauge value at fixed lattice spacing 
corresponding to
$6/(g_E^2 a)=21$. The analysis of Ref.~\cite{hart} also suggests that a possible
temperature dependence of $c$ is less than $5\%$. Therefore we use an averaged value
$c=0.54(1)$ in our analysis, where the indicated uncertainty also includes possible temperature
dependence in it.

\begin{figure}
\includegraphics[width=9.0cm]{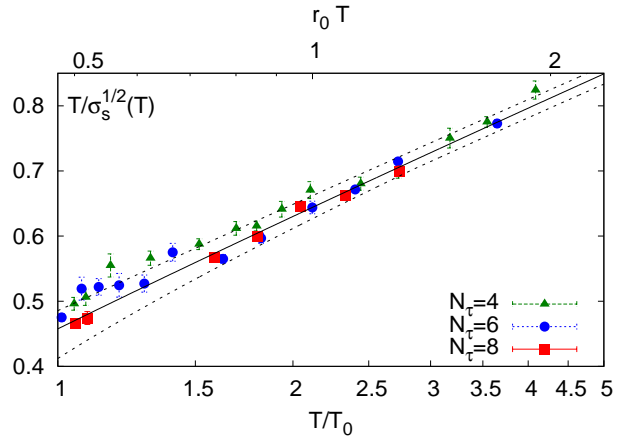}
\caption{$T/\sqrt{\sigma_s(T)}$ calculated on $N_{\tau}=4,~6$ and $8$ lattices
compared with the prediction of dimensional reduction indicated by the line. 
The uncertainty in the prediction of the dimensionally reduced theory is shown
by dashed lines. For easier visualization the temperature on the 
lower horizontal axis 
has been rescaled by $T_0=200$ MeV corresponding to $r_0 T_0=0.47619$.
We also show the temperature scale in units of $r_0$ on the upper 
horizontal axis.
}
\label{fig:comp3d}
\end{figure}

The gauge coupling of the effective 3d theory has been calculated to
2-loop accuracy \cite{mikko}
\begin{eqnarray}
\frac{g_E^2(T,\bar \mu)}{T}&=&
g^2(\bar \mu)+g^4(\bar \mu) 
\left( 2 b_0 \ln\left(\frac{\bar \mu}{\pi T}\right)+a_2 \right)+\nonumber\\
&&
g^6(\bar \mu) \left( 2 b_1 \ln \left(\frac{\bar \mu}{\pi T} \right)\right.+
\nonumber \\
&&\left. \left[2 b_0 \ln \left(\frac{\bar \mu}{\pi T} \right) -
\frac{8 n_f \ln 2 -3}{48 \pi^2 }\right]^2 +a_3\right) .
\label{ge2}
\end{eqnarray}
Here $g(\bar \mu)$ is the QCD coupling in  $\overline{MS}$ scheme, $b_0=(11-2 n_f/3)/(4 \pi)^2$
and $b_1=(102-38 n_f/3)/(4 \pi)^4$ are the coefficients of the universal 2-loop
beta function, and $\bar \mu$ is the $\overline{MS}$ renormalization scale.
The coefficients $a_2$ and $a_3$ can be found in Ref. \cite{mikko}.
In our case $n_f=3$. The coupling $g_E^2$ depends on the renormalization
scale $\bar \mu$ at any fixed order of perturbation theory. Of course, the
dependence on $\bar \mu$ gets weaker and weaker as we go to higher orders
of the perturbative expansion. In practice, however, we have to deal with the
scale dependence of the effective coupling. Following Ref. \cite{mikko} 
we fix the scale $\bar \mu_*$  using the principle of minimum sensitivity, i.e. 
we require that the derivative of the 1-loop expression 
for $g_E^2(T,\bar \mu)$ vanishes at $\bar \mu=\bar \mu_*$,   
and vary the scale in the interval $\bar \mu=(0.5-2)\bar \mu_*$. 
For 3-flavor QCD we find the value $\bar \mu_* \simeq 9.1T$.
To specify $g(\bar \mu)$ and thus $g_E$ completely we need to know the value 
of $\Lambda_{\overline{MS}}$, or more
precisely the ratio $T/\Lambda_{\overline{MS}}$.
Since the temperature has been set by the Sommer parameter $r_0$ 
this means that we need to specify 
$\Lambda_{\overline{MS}} r_0$.
This can be done in principle by calculating the $r_0$ parameter 
at several lattice spacings
and fitting it by the modified 2-loop Ansatz \cite{alton} 
to determine $r_0 \Lambda_{Lat}$.
One then may use lattice perturbation theory to calculate 
$\Lambda_{\overline{MS}}/\Lambda_{Lat}$.
This has been done in SU(3) gauge theory \cite{capitani} 
as well as in 2-flavor QCD
with Wilson fermions \cite{goeckeler}. Unfortunately, the 
perturbative calculations needed for this have
not been performed for the p4fat3 action. On the other hand we 
can express the temperature in physical units using the Sommer parameters
$r_0=0.469(7)$fm obtained from quarkonium splitting as input \cite{gray}.  
Furthermore, using the same physical input 
the running coupling constant has been calculated on the lattice in 2+1 flavor 
QCD within the so-called
$V$ scheme \cite{mason}, giving $\alpha_V(\mu=7.5GeV)=0.2082(40)$. 
This corresponds to 
$\alpha_{\overline{MS}}(\bar \mu=7.5GeV)=g^2(\bar \mu)/(4 \pi)=0.180(+2)(-3)$ 
if we use the 3-loop relation \cite{york} between the coupling in 
the $V$-scheme and the $\overline{MS}$-scheme.  
Using the 2-loop beta function we can determine the
coupling $g(\bar \mu)$ entering Eq. (\ref{ge2}) at the given scale 
$\bar \mu_*$ thus specifying 
the gauge coupling of the effective 3d theory
$g_E^2(T)$ for different temperatures. Combining the value of $c$ 
from above and $g_E^2(T)$ we finally get
the corresponding prediction for the spatial string tension in the
dimensionally reduced theory. 
In Figure \ref{fig:comp3d} we show our results for $T/\sqrt{\sigma_s(T)}$ 
compared with the prediction coming from the dimensionally reduced theory
as discussed above. This is  
shown as a solid line with its uncertainty shown as a band (dashed lines).
There are three sources of  uncertainty in the dimensional reduction prediction.
The first is the uncertainty in the value of $c$. 
This is the dominant source of the
error at high temperatures, $T>2 T_c$. 
The second source of error is the scale dependence
of $g_E$, which is the most important one at low temperatures, $T<2 T_c$. 
Finally there is an error
in $g(\bar \mu)$ coming from the value of  
$\alpha_{\overline{MS}}(7.5GeV)=0.180(+2)(-3)$. This, however,
is significantly smaller than the previous two in the entire temperature range.

Figure \ref{fig:comp3d} seems to suggest 
that dimensional reduction works down to temperatures surprisingly close to 
the transition temperature. 
In order to see whether this 
picture is self-consistent one has to
calculate other spatial correlation functions.
In particular, one has to verify that the largest correlation length
of operators built from quarks is sufficiently small to justify
integrating out the Matsubara modes of quarks.
In Ref.~\cite{gavai} this problem 
has been studied in 4-flavor QCD,
using $N_{\tau}=4$ lattices with the standard staggered formulation. 
The analysis performed in that paper suggests that the pion
correlator gives the largest correlation length up to temperatures 
as high as $3T_c$.
Note, however, that this may be due to large discretization errors 
in the screening masses 
when the standard staggered action is used on
$N_{\tau}=4$ lattices. 
Indeed, for this action
it has been noticed in Ref.~\cite{gavai03} that the pion screening
masses on $N_{\tau}=10-12$ 
lattices may come close to a value of $2 \pi T$.
Recent preliminary calculations with improved (p4fat3) staggered fermions
\cite{mukherjee,mscr_unpub} give large values 
for this quantity,
$m_{\pi}/T \simeq 4.6$ for $T \simeq 1.5T_c$
and 
$m_{\pi}/T \simeq 5$ for $T \simeq 2.0T_c$,
already on lattices with $N_{\tau}=4$ and $6$.
Thus the value of the pion screening mass is larger than 
the smallest glueball screening mass 
for $T \ge 1.5T_c$ which was estimated to be 
$\sim 4 T$ \cite{gavai}
and larger than the Debye screening mass estimated in \cite{mD2+1f}. 
This indeed would suggest that 
in QCD dimensional reduction 
may
work down to temperatures as low
as $1.5T_c$ also in the case of 2+1 flavor.   

\section{Conclusions}
In this paper we have calculated the spatial string tension in 2+1 flavor QCD
with a physical strange quark mass and light quark masses of $0.1m_s$ 
corresponding to a pion mass of about $220\;$MeV. 
The spatial string tension calculated at different
lattice spacings agree reasonably well with each other. 
We have compared the results
of our calculation with the prediction of dimensionally reduced effective 
theory and have found remarkably good agreement down to temperatures close 
to the transition temperature. This is similar to the observation made 
in SU(3) gauge theory \cite{mikko}.
There are three sources of uncertainty when comparing the data on the 
spatial string tension the scale dependence of the 3d gauge coupling, 
the uncertainty in the value of coefficient $c$
and the uncertainty in $r_0 \Lambda_{\overline{MS}}$. 
The uncertainties from the last two sources could be reduced by calculating 
the lattice
beta function for the p4fat3 action perturbatively and through a more 
precise calculation of the string tension
of the 3d adjoint Higgs model. 

Let us finally note that the spatial string tension has been recently
studied also in 2 flavor QCD using Wilson fermions and significantly 
larger quark masses \cite{bornyakov,maezawa}.
In Ref. \cite{bornyakov} the spatial string 
tension has been calculated only up to $1.28T_c$
and no temperature dependence has been found in this temperature interval.
The results of Ref. \cite{maezawa}, obtained on lattices with temporal extent $N_{\tau}=4$ 
qualitatively agree with ours,  but the drop of the string tension close to $T_c$
is significantly larger.  
It remains to be seen
to what extent these discrepancies are due to larger quark mass, 
cutoff effects or limited statistics, as 
calculations with Wilson fermions are numerically more demanding.

\section*{Acknowledgements}
EL and JL wish to thank Y. Schr\"oder for helpful discussions.
This work has been supported in part by contracts DE-AC02-98CH1-0886 
and DE-FG02-92ER40699 with the U.S. Department of Energy, 
by the Bundesmi\-ni\-sterium f\"ur Bildung und Forschung
under grant 06BI401, the Gesellschaft f\"ur Schwerionenforschung 
by contract BILAER
and the Deutsche Forschungsgemeinschaft under grant GRK 881. 
Calculations reported here were carried out using
the QCDOC supercomputers of the RIKEN-BNL Research Center and the U.S.
DOE, apeNEXT supercomputer
at Bielefeld University, and the BlueGene/L (NYBlue) at the New York Center for 
Computational Sciences at Stony Brook University/Brookhaven 
National Laboratory which is supported by the U.S. Department 
of Energy under Contract No. DE-AC02-98CH10886 and by the 
State of New York.

\end{document}